%% file: ProductRuleQMetro.tex
\documentclass[12pt]{iopart}

\usepackage{iopams}  
\usepackage{xypic}

\def\Comb{\mathsf{Comb}}

\def\Tr{{\rm Tr}} 
\def\St{{\sf St}}

\def\spc#1{\mathcal{#1}}
\def\map#1{\mathcal{#1}}
\def\>{\rangle}
\def\<{\langle}
\def\qed{$\blacksquare$}
\def\Lin{\mathsf{Lin}}
\def\Cmplx{\mathbb{C}}
\def\Reals{\mathbb{R}}

\def\Proof{{\bf  Proof.}~}
\def\d{{\rm d}}

\newtheorem{theo}{Theorem}
\newtheorem{cor}{Corollary}

\input{myQcircuit}
\begin{document}

\title{
Optimal networks for Quantum Metrology: semidefinite programs and product rules 
}
\author{Giulio Chiribella} 
\address{Center for Quantum Information, Institute for Interdisciplinary Information Sciences, Tsinghua University, Beijing 100084, China. }

\begin{abstract}
We investigate the optimal estimation of  a quantum process  that can possibly consist of multiple time steps.    The estimation is implemented by a quantum network that interacts with the process by sending an input and processing the output at each time step. 
We formulate the search of the optimal  network as a semidefinite program and use duality theory to give an alternative expression for the maximum payoff achieved by estimation.  Combining this formulation with a technique devised by Mittal and Szegedy we prove a general product rule for the joint estimation of independent processes, stating that the optimal  joint estimation can  achieved by estimating each process independently, whenever the figure of merit is of a product form.   
We illustrate the result in several examples and exhibit counterexamples showing that the optimal joint network may not be the product of the optimal individual networks if the processes are not independent or if the figure of merit is not of the  product form.  In particular, we show that  entanglement can reduce by a factor $K$ the variance in the estimation of the sum of $K$ independent phase shifts.  
 \end{abstract}

\maketitle

\section{Introduction}

Quantum theory offers impressive advantages over classical theory in the estimation of  physical parameters \cite{caves,wineland,derka,phase,direction,peres,refframe,qmetro1,giova,phasenet,qmetro2}.  The prototypical example is  the estimation of an unknown phase shift \cite{derka,phase,giova,phasenet}: here the variance vanishes as $N^{-2}$ with the number $N$ of accesses to the phase-shifting process, whereas a classical statistics over independent copies would give the scaling $N^{-1}$.   The quadratic improvement is achieved by preparing an entangled state of $N$ systems and applying the unknown process to each system.  The same quadratic advantage can be found in the estimation of a direction in space \cite{direction,peres} and in the joint estimation of three Cartesian axes \cite{refframe,baganis,haypla}.

Given the usefulness of entanglement in the estimation of a single parameter from multiple accesses to a physical process,   
it is natural to ask whether entanglement can improve the estimation of many  parameters corresponding to different processes. 
For example, one may wonder whether entanglement can help in the estimation of two independent phase shifts.  In a slightly different context, this type of question was originally addressed by Wootters in an unpublished work and by DiVincenzo, Terhal, and Leung  \cite{woot}, who asked whether a joint entangled measurement could improve the extraction of information about two bits encoded in two independent sets of states. In this scenario, it was shown that that the amount of information that can be extracted from the product set is additive  \cite{woot}.  More recently, a different proof showing the optimality of product measurements for the extraction of information from general product sets of states was provided in  Ref. \cite{sinoprod}. 

In this paper we address the problem of the joint estimation of the parameters encoded in a set of independent processes, where each process can consist of several time steps.  Due to the possibility of connecting an input of an unknown process with the output of another one, here the question whether quantum correlations can improve the estimation  is not only a question about the usefulness of entanglement in the input states and in the measurements, but also a question about  the usefulness of quantum correlations \emph{in time}, namely correlations mediated by the exchange of quantum systems from one time step to the next.      
We address the question in the framework of quantum estimation \cite{hel,hol},  where the figure of merit is the expected payoff associated to a payoff function $g (\hat x, x)$, which depends of the true value $x$ and of estimated value $\hat x$ labelling the unknown process.   In order to tackle the question we  formulate the optimization of the quantum network for the estimation of an unknown multi-time process as a semidefinite program and we discuss the corresponding dual problem.   
In this context we prove a general product rule, showing that the optimal joint estimation of a set of independent parameters ${\bf x}  := (x_1, \dots, x_K)$  can be achieved by estimating each parameter independently whenever the figure of merit if of the product form    $g (\hat {\bf x},  {\bf  x})  =\prod_{k=1}^K  g_k (\hat x_k, x_k) $, where $ g_k$ is the payoff function for the parameter $x_k$.   In particular, our result implies that the maximum probability of success in identifying a set of unknown processes is the product of the maximum probabilities of success in identifying each individual process separately.

Product theorems are a key tool in theoretical computer science  \cite{feige,raz,holstein,cleve,tlee,mittszeg,lee}, where one is often interested in how the resources needed to solve several independent problems jointly are related to the resources needed to solve each problem individually.  Our work begins to explore the usefulness of this techniques in the domain of physics, starting from the fundamental problem of identifying a set of  independent physical parameters.     
In order to prove our result we use the framework of \emph{quantum combs} \cite{combprl,combpra} (see also the work by Gutoski and Watrous on \emph{quantum strategies} \cite{watgut}). As we already mentioned, in this framework we  formulate the maximization of the expected payoff as a semidefinite program, and present an intuitive formulation of the dual minimization program.    Such a dual formulation is interesting in its own right, as it generalizes to arbitrary processes and arbitrary payoff functions a classic formula derived  by Yuen, Kennedy, and Lax \cite{yuen}   for the  minimum error state discrimination. 
Exploiting the form of the primal and dual programs, we then prove our product theorem following a general technique devised by Mittal and Szegedy in Ref. \cite{mittszeg} (see also Ref. \cite{lee}), which is adapted here in order to deal with the optimization of quantum networks consisting of multiple time steps.

\section{Quantum networks for process estimation}\label{sec:combs}  

Suppose that  an experimenter has access to a physical process $\map P_x$ that depends on  an unknown parameter $x$ in some parameter space $\mathsf X$.  The goal of the experimenter is  to determine the parameter $x$ with the maximum precision allowed by the laws of quantum mechanics.    

Generally, the process $\map  P_x$ can consist of $N$ time steps, labelled by an index $s$ in some finite set $\mathsf S   =  (s_1, \dots s_N) \subset \mathbb N$, ordered so that $s_m  <  s_n$ for $m<  n$.     At each time step $s \in \mathsf S $    the process transforms an input quantum system, with Hilbert space denoted by $\spc H_{in}^{(s)}$, into a  (possibly different) output quantum system,  with Hilbert space denoted by $\spc H_{out}^{(s)} $.     
If the process   $\map P_x $ is memoryless, all time steps are independent and one can associate a quantum channel to each time step. The quantum channel  at step  $s$, denoted by $\map C^{(s)}_x $, will be a completely positive trace-preserving map sending density matrices on $\spc H_{in}^{(s)}$ to density matrices on $\spc H_{out}^{(s)}$.   Hence, the process $\map P_x$ can be described by a time-ordered sequence of quantum channels, each channel labelled by the unknown parameter $x$, as in the following picture:  

\begin{equation*}
 \Qcircuit @C=1em @R=.7em @! R {
&\qw \poloFantasmaCn{\spc H_{in}^{(s_1)}}   &\gate{\map C^{(1)}_x }  & \qw \poloFantasmaCn{\spc H_{out}^{(s_1)}}  & \qw &  & & \qw\poloFantasmaCn{\spc H_{in}^{(s_2)}}  & \gate{\map C^{(2)}_x}    &  \qw \poloFantasmaCn{\spc H_{out}^{(s_2)}}   &\qw &  &  & \poloFantasmaCn{\dots\quad } &  &  &\qw &     \qw \poloFantasmaCn{\spc H_{in}^{(s_N)}}  &\qw &   \gate{\map C^{(N)}_{x}}  &\qw &   \qw \poloFantasmaCn {\spc H_{out}^{(s_N)}}  &  \qw   &} 
 \end{equation*}  
In the easiest case, one may have the same channel at each time step, namely $\map C_x^{(s)} =   \map C_x$ for every $s\in \mathsf S$.  This is the case, e.g. of quantum phase estimation \cite{derka,phase,qmetro1,giova,phasenet,qmetro2}, where one has access to $N$ uses of the unitary channel $\map C_x  =  U_x \rho  U_x^\dag$,  with $U_x  =  \exp( i x H)$ for some Hamiltonian $H$ with integer spectrum.

In the presence of memory, the input-output transformation at the step $s$ is described by a quantum channel involving internal ancillas: in this case  the quantum channel $\map C_x^{(s)}$  transforms density matrices on   $\spc H_{in}^{(s)}  \otimes \spc A_{s-1}$ to density matrices on $\spc H_{out}^{(s)}  \otimes \spc A_{s}$, where $\spc A_s$ is the Hilbert space of the $s$-th ancilla. Hence, the process $\map C_x$ is represented by a time-ordered sequence of black boxes with internal memories:   

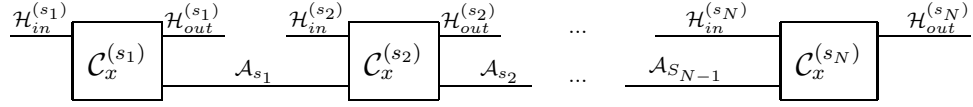
\begin{equation*} 
 \Qcircuit @C=1em @R=.7em @! R {
&\qw \poloFantasmaCn{\spc H_{in}^{(s_1)}}   &\multigate{1}{\map C^{(s_1)}_x }  & \qw \poloFantasmaCn{\spc H_{out}^{(s_1)}}  & \qw &  & & \qw\poloFantasmaCn{\spc H_{in}^{(s_2)}}  & \multigate{1}{\map C^{(s_2)}_x}    &  \qw \poloFantasmaCn{\spc H_{out}^{(s_2)}}   &\qw &  &  & \poloFantasmaCn{\dots\quad } &  &  &\qw &     \qw \poloFantasmaCn{\spc H_{in}^{(s_N)}}  &\qw &   \multigate{1}{\map C^{(s_N)}_{x}}  &\qw &   \qw \poloFantasmaCn {\spc H_{out}^{(s_N)}}  &  \qw   &\\
&    &\pureghost{\map C^{(s_1)}_x }  & \qw   & \qw &  \qw \poloFantasmaCn{\spc A_{s_1}}   & \qw  & \qw  & \ghost{\map C^{(s_2)}_x}    &  \qw    &\qw \poloFantasmaCn{\spc A_{s_2}} & \qw  &  & \poloFantasmaCn{\dots\quad } &  & \qw &\qw \poloFantasmaCn{\spc A_{S_{N-1}}} &     \qw   &\qw &   \ghost{\map C^{(s_N)}_{x}}  &  &     &    &} 
 \end{equation*}

Note that, since the ancillas are internal to the network, the first and last ancillary systems are trivial $\spc A_0  \simeq \spc A_{N}  \simeq \mathbb C$.      

The most general strategy to estimate an unknown parameter from a time-ordered sequence of black boxes consists in inserting them in a quantum network where they are interspersed with known quantum gates and eventually a quantum measurement is performed on the output, producing the estimate $\hat x\in \mathsf X$.  

The estimation process can be depicted as 
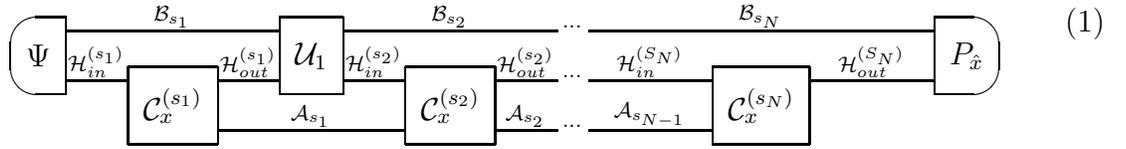
\begin{equation}\label{testedwithmemo}  
\mbox{\Qcircuit @C=1em @R=.7em @! R {
\pureghost{\Psi} &\qw &\qw  \poloFantasmaCn{\spc B_{s_1}}  & \qw & \ghost{\map U_{1}}  &  \qw  & \qw   \poloFantasmaCn{\spc B_{s_2}} & \qw &\qw  &\poloFantasmaCn{ \dots\quad } & \qw & \qw & \qw  &\qw   \poloFantasmaCn{\spc B_{s_N}}& \qw &\qw &\qw&\ghost{P_{\hat x}} \\
\multiprepareC{-1}{   \Psi} &\qw \poloFantasmaCn{\spc H^{(s_1)}_{in}}   &\multigate{1}{\map C^{(s_1)}_x }  & \qw \poloFantasmaCn{\spc H^{(s_1)}_{out} }  & \multigate{-1}{\map U_1} & \qw\poloFantasmaCn{\spc H_{in}^{(s_2)}}  & \multigate{1}{\map C^{(s_2)}_x}    &  \qw \poloFantasmaCn{\spc H_{out}^{(s_2)}}   & \qw &   \poloFantasmaCn{\dots\quad } &  \qw &     \qw \poloFantasmaCn{\spc H_{in}^{(S_N)}}  &\qw &   \multigate{1}{\map C^{(s_N)}_{x}}  &\qw &   \qw \poloFantasmaCn {\spc H_{out}^{(S_N)}}  &  \qw   &\multimeasureD{-1}{P_{\hat x}}  
\\
  &    &\pureghost{\map C^{(s_1)}_x }  & \qw  & \qw \poloFantasmaCn{\spc A_{s_1}}    & \qw & \ghost{\map C^{(s_2)}_x}    &  \qw \poloFantasmaCn{\spc A_{s_2}}   & \qw &   \poloFantasmaCn{\dots\quad } &  \qw &     \qw \poloFantasmaCn{\spc A_{s_{N-1} }}  &\qw &   \ghost{\map C^{(s_N)}_{x}}  &  &   &     &}    }     \qquad
 \end{equation}  
 
\noindent 
where $\spc B_s,~ s\in\mathsf S $ are the internal ancillas of the estimating network, $\Psi$  is a quantum state on $ \spc B_{s_1} \otimes \spc H^{(s_1)}_{in}$, each $\map U_s$ is a quantum channel, and $P_{\hat x}$ is a quantum measurement, described by a \emph{positive operator valued measure (POVM)} on the  Hilbert space  $\spc B_{s_N} \otimes \spc H^{(s_N)}_{out}$. 

Examples of quantum networks for the estimation of unknown parameters can be found in Refs. \cite{giova,phasenet}.

\section{Optimizing quantum networks: the method of quantum combs}

A convenient way to optimize quantum networks is the method of \emph{quantum combs} \cite{combprl,combpra} (see also the work on \emph{quantum strategies} by Gutoski and Watrous \cite{watgut}), which associates positive operators to sequential quantum networks.  Here we briefly summarize some known basic facts about this method, referring the reader to the original papers for the proofs and for further details.

In the following we will use the following notation:   $\Lin (\spc H)$ will denote the set of linear operators on a (finite-dimensional)  Hilbert space $\spc H$,   $\Lin_+(\spc H)$  will denote the set of positive operators on  $\spc H$, while $\St (\spc H)$ will denote the set of density matrices on $\spc H$, that is, the set of positive operators $\rho\in \Lin_+ (\spc H)$ such that $\Tr[\rho]=1$.

\subsection{Quantum combs.}  
A sequential network of quantum channels with internal memories can be associated with a non-negative operator satisfying suitable linear constraints. Precisely, a network of the form

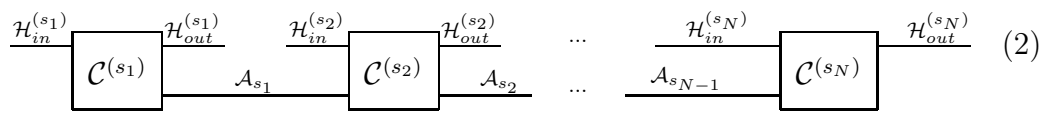
\begin{equation}\label{withmemo}  
  \Qcircuit @C=1em @R=.7em @! R {
&\qw \poloFantasmaCn{\spc H_{in}^{(s_1)}}   &\multigate{1}{\map C^{(s_1)} }  & \qw \poloFantasmaCn{\spc H_{out}^{(s_1)}}  & \qw &  & & \qw\poloFantasmaCn{\spc H_{in}^{(s_2)}}  & \multigate{1}{\map C^{(s_2)}}    &  \qw \poloFantasmaCn{\spc H_{out}^{(s_2)}}   &\qw &  &  & \poloFantasmaCn{\dots\quad } &  &  &\qw &     \qw \poloFantasmaCn{\spc H_{in}^{(s_N)}}  &\qw &   \multigate{1}{\map C^{(s_N)} }  &\qw &   \qw \poloFantasmaCn {\spc H_{out}^{(s_N)}}  &  \qw   &\\
  &  &\pureghost{\map C^{(s_1)}  }  & \qw   & \qw &  \qw \poloFantasmaCn{\spc A_{s_1}}   & \qw  & \qw  & \ghost{\map C^{(s_2)} }    &  \qw    &\qw \poloFantasmaCn{\spc A_{s_2}} & \qw  &  & \poloFantasmaCn{\dots\quad } &  & \qw &\qw \poloFantasmaCn{\spc A_{s_{N-1}}} &     \qw   &\qw &   \ghost{\map C^{(s_N)}}  &  &     &    &} 
 \end{equation}  
is associated  with a positive operator $R  \in  \Lin_+  \left [  \bigotimes_{s\in \mathsf S}   \left (   \spc H_{out}^{(s)} \otimes \spc H_{in}^{(s)}  \right)  \right]$. The fact that the network consists of quantum channels (trace-preserving maps) imposes the following constraint:  there must exist a set of positive operators  $R^{( n )}  \in   \Lin_+   \left [  \bigotimes_{i =1}^n   \left (   \spc H_{out}^{(s_i)} \otimes \spc H_{in}^{(s_i)}  \right)  \right]$, $ n=1, \dots , N-1$  such that 
\begin{equation}\label{normcomb}
\left\{  
\begin{array}{rcl} 
 \Tr_{out, s_N}  \left[R \right]   &=&  I_{in, s_N}  \otimes R^{(N-1)}\\    
\Tr_{out, s_{N-1}}  \left[R^{(N-1)}\right]   &=&  I_{in,s_{N-1}}  \otimes R^{(N-2)}\\
 &  \vdots &    \\
\Tr_{out, s_1}  \left[R^{(1)}\right]   &=&  I_{in,s_1} ,
\end{array}
\right.
\end{equation} 
where $\Tr_{out, s}$ and $I_{in,s}$ denote the partial trace over $\spc H^{(s)}_{out}$ and the identity operator on $\spc H_{in}^{(s)}$, respectively  \cite{watgut,combprl,combpra}.  
 
 Most importantly, the converse also holds \cite{watgut,combprl,combpra}: if a positive operator $R$   satisfies the constraints of Eq. (\ref{normcomb}) for some set of positive operators $R^{(n)}, n=1, \dots , N-1$, then there exists a network of the form of Eq. (\ref{withmemo}) such that the operator associated to that network is $R$.  This is important because it implies that optimizing over quantum networks is completely equivalent to optimizing over positive operators $R$ satisfying Eq. (\ref{normcomb}).        In fact, given an operator $R$ satisfying there is a constructive algorithm to build up the channels $\map C^{(s)}$ at all time steps $s\in\mathsf S$  \cite{bisio}. In the following, a positive  operator $R  \in  \Lin_+  \left [  \bigotimes_{s\in \mathsf S}   \left (   \spc H_{out}^{(s)} \otimes \spc H_{in}^{(s)}  \right)  \right]$   satisfying Eq. (\ref{normcomb})  for some operators $R^{(n)}, n=1, \dots , N-1$ will be called  \emph{quantum comb}.   We will denote the set of quantum combs with a prescribed number of time steps and prescribed input and output Hilbert spaces as $  \Comb   \left[  \bigotimes_{s\in \mathsf S}   \left (   \spc H_{out}^{(s)} \otimes \spc H_{in}^{(s)}  \right)  \right]$.   

\subsection{Quantum testers.} 

More generally, a quantum network can contain measurements: at each time step $s$   one can have  a   measurement with outcome $m_s $ in some set $ \mathsf M_s$.  
Conditionally to the outcome $m_s$, the input system  will undergo a  transformation, represented by a completely positive trace non-increasing map  $  \map C^{(s)}_{m_s}$, with the condition that the sum over all outcomes  $  \map C^{(s)}  :=  \sum_{m_s\in\mathsf M_s}\map C^{(s)}_{m_s}$ is trace-preserving.       
A sequential network containing measurements, such as the network  

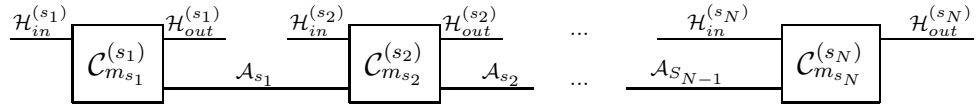
\begin{equation*} 
  \Qcircuit @C=1em @R=.7em @! R {
&\qw \poloFantasmaCn{\spc H_{in}^{(s_1)}}   &\multigate{1}{\map C^{(s_1)}_{m_{s_1}} }  & \qw \poloFantasmaCn{\spc H_{out}^{(s_1)}}  & \qw &  & & \qw\poloFantasmaCn{\spc H_{in}^{(s_2)}}  & \multigate{1}{\map C^{(s_2)}_{m_{s_2}}}    &  \qw \poloFantasmaCn{\spc H_{out}^{(s_2)}}   &\qw &  &  & \poloFantasmaCn{\dots\quad } &  &  &\qw &     \qw \poloFantasmaCn{\spc H_{in}^{(s_N)}}  &\qw &   \multigate{1}{\map C^{(s_N)}_{m_{s_N}}}  &\qw &   \qw \poloFantasmaCn {\spc H_{out}^{(s_N)}}  &  \qw   &\\
&    &\pureghost{\map C^{(s_1)}_{m_{s_1}} }  & \qw   & \qw &  \qw \poloFantasmaCn{\spc A_{s_1}}   & \qw  & \qw  & \ghost{\map C^{(s_2)}_{m_{s_2}}}    &  \qw    &\qw \poloFantasmaCn{\spc A_{s_2}} & \qw  &  & \poloFantasmaCn{\dots\quad } &  & \qw &\qw \poloFantasmaCn{\spc A_{S_{N-1}}} &     \qw   &\qw &   \ghost{\map C^{(s_N)}_{m_{s_N}}}  &  &     &    &} 
 \end{equation*}  
can be associated with a collection of positive operators ${\bf T}   :=  \{T_m~|~  m  \in  \mathsf M  :  =\mathsf M_1  \times \dots \times \mathsf M_N\}$ with the property that the sum over all outcomes $T:  =  \sum_{m \in \mathsf M} T_m $ satisfies Eq. (\ref{normcomb}).   We call such a collection of operators   a \emph{quantum tester}. It is possible to prove that, if a collection positive operators ${\bf T}  =  \{T_m~|~  m  \in  \mathsf M  \}$  is a quantum tester, then there exists a quantum network of the form  
   
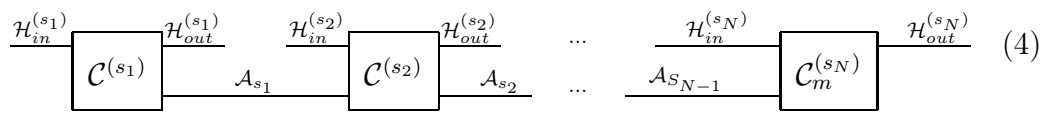
\begin{equation}\label{testerpicu}  
  \Qcircuit @C=1em @R=.7em @! R {
&\qw \poloFantasmaCn{\spc H_{in}^{(s_1)}}   &\multigate{1}{\map C^{(s_1)} }  & \qw \poloFantasmaCn{\spc H_{out}^{(s_1)}}  & \qw &  & & \qw\poloFantasmaCn{\spc H_{in}^{(s_2)}}  & \multigate{1}{\map C^{(s_2) } }     &  \qw \poloFantasmaCn{\spc H_{out}^{(s_2)}}   &\qw &  &  & \poloFantasmaCn{\dots\quad } &  &  &\qw &     \qw \poloFantasmaCn{\spc H_{in}^{(s_N)}}  &\qw &   \multigate{1}{\map C^{(s_N)}_{m}}  &\qw &   \qw \poloFantasmaCn {\spc H_{out}^{(s_N)}}  &  \qw   &\\
&    &\pureghost{\map C^{(s_1)} }  & \qw   & \qw &  \qw \poloFantasmaCn{\spc A_{s_1}}   & \qw  & \qw  & \ghost{\map C^{(s_2)}}    &  \qw    &\qw \poloFantasmaCn{\spc A_{s_2}} & \qw  &  & \poloFantasmaCn{\dots\quad } &  & \qw &\qw \poloFantasmaCn{\spc A_{S_{N-1}}} &     \qw   &\qw &   \ghost{\map C^{(s_N)}_{m}}  &  &     &    &} 
 \end{equation}  
such that $\mathbf T$ is the tester associated to that network \cite{watgut,combprl,combpra}.   Note that here the measurement takes place only in the last step, while the boxes $\map C^{(s_n)}$, $n=1.\dots, N-1$ represent quantum channels.   

A particular type of testers are those where the first and last quantum systems are trivial [$\spc H_{in}^{(s_1)}  \simeq  \spc H_{out}^{(s_N)}  \simeq \Cmplx$  in Eq. (\ref{testerpicu})]. These testers represent quantum networks that start with a state preparation and end with a POVM measurement.     These are exactly the networks that are interesting for the estimation of quantum processes, as depicted in Eq. (\ref{testedwithmemo}): note that to test a process consisting of $N$ time steps we need  tester consisting of $N+1$ time steps. Labelling the Hilbert spaces as in the following diagram

\begin{equation}\label{testerpic}  
  \Qcircuit @C=1em @R=.7em @! R {
   &\multiprepareC{1}{\map C^{(s_1)} }  & \qw \poloFantasmaCn{\spc H_{in}^{(s_1)}}  & \qw &  & & \qw\poloFantasmaCn{\spc H_{out}^{(s_1)}}  & \multigate{1}{\map C^{(s_2) } }     &  \qw \poloFantasmaCn{\spc H_{in}^{(s_2)}}   &\qw &  &  & \poloFantasmaCn{\dots\quad } &  &  &\qw &     \qw \poloFantasmaCn{\spc H_{out}^{(s_N)}}  &\qw &   \multimeasureD{1}{\map C^{(s_{N+1})}_{m}}  \\
 &\pureghost{\map C^{(s_1)} }  & \qw   & \qw &  \qw \poloFantasmaCn{\spc A_{s_1}}   & \qw  & \qw  & \ghost{\map C^{(s_2)}}    &  \qw    &\qw \poloFantasmaCn{\spc A_{s_2}} & \qw  &  & \poloFantasmaCn{\dots\quad } &  & \qw &\qw \poloFantasmaCn{\spc A_{S_{N}}} &     \qw   &\qw &   \ghost{\map C^{(s_{N+1})}_{m}}  } 
 \end{equation}  
the normalization of the tester $\mathbf T$ becomes  
\begin{equation}\label{testernorm}
\left\{  
\begin{array}{rcl}
\sum_{m\in\mathsf M}  T_m   & =&   I_{out,s_N}  \otimes \Xi^{(N)}\\
\Tr_{in,s_N} \left[\Xi^{(N)}\right]  & = & I_{out,s_{N-1}}  \otimes \Xi^{(N-1)}\\
  & \vdots  & \\
\Tr_{in,s_1} [ \Xi^{(1)}]& =&  1,
\end{array} 
\right.
\end{equation}
for some set of positive operators  $\Xi^{(n)}\in \Lin \left\{  \spc H_{ in}^{(s_n) } \otimes \left[ \bigotimes_{i=1}^{ n-1} \left(  \spc H_{out}^{(s_i)}  \otimes \spc H_{in}^{(s_i)}  \right)\right] \right\}$, $n=1,\dots,N$. 

\subsection{Generalized Born rule.}  
 If we test a process represented by the quantum comb $R \in     \Comb   \left[  \bigotimes_{s\in \mathsf S}   \left (   \spc H_{out}^{(s)} \otimes \spc H_{in}^{(s)}  \right)  \right]$  with a network represented by the tester $\mathbf T := \{   T_m ~|~  m\in \mathsf M \}$, then we obtain a probability distribution $p(m|  R^{(N)})$ over all possible outcomes. Such a probability distribution is given by the \emph{generalized Born rule} of Refs.  \cite{combprl,combpra}:
\begin{equation}\label{born}
p\left(m| R  \right)  =    \Tr\left[    T_ m  R   \right].  
\end{equation}
Here the quantum comb $R$ plays the role of the density matrix in the ordinary Born rule, and the tester   $\{   T_m ~|~  m\in \mathsf M \}$ plays the role of the POVM measurement.    In fact, the ordinary Born rule can be retrieved as a special case of Eq. (\ref{born}), corresponding to the case of \emph{state preparation processes},  namely processes that consist of a single time step  ($N=1$) with no input system ($\spc H_{in}^{(s_1)}  \simeq \Cmplx$). In that special case, the normalization of the quantum comb, given by  $\Tr_{out, s_1} [R]  =  I_{in, s_1}$ becomes $\Tr[R]  = 1$, which is the normalization of a density matrix, while the normalization of the tester, given by $\sum_{m \in\mathsf M}   T_m  =   I_{out, s_1} \otimes \Xi^{(1)}$,  $\Tr[\Xi^{(1)}]  =1$,  becomes $\sum_{m \in\mathsf M}   T_m  =   I_{out, s_1}$, which is the normalization of a POVM.


\section{The optimization problem of Quantum Metrology}  

In process estimation one has a parametric family of processes with a given input-output structure and with a fixed number of time steps $N$ labelled by an index $s \in\mathsf S  \subset \mathbb N$ .  Each process is described by a quantum comb $R_x \in  \Comb   \left[  \bigotimes_{s\in \mathsf S}   \left (   \spc H_{out}^{(s)} \otimes \spc H_{in}^{(s)}  \right)  \right]$, where $x \in  \mathsf X$ is the  parameter to be estimated.   Let us denote by $\pi(x)$  the probability that the unknown parameter has the value $x$.  
 If $x$ has a continuum of values, $p(x)$  will represent the probability density of $x$ with  respect to some measure $\d x$.   For simplicity in the following we will present the results in the discrete case, but it is important to bear in mind that these results hold also in the continuous case, just replacing sums with integrals and replacing the quantifier ``$\forall \hat x  \in  \mathsf X$" with ``$\forall \hat x  \in  \mathsf X$ except at most for a set of zero measure".

\subsection{Primal maximization problem}  
 For an estimation strategy  described by the quantum tester $\mathbf T  :=\{  T_{\hat x}  ~|~  \hat x  \in \mathsf X\}$,  the probability distribution $p(\hat x | x)$ is given by Eq. (\ref{born}). In order to evaluate the performance of a given strategy, we introduce a payoff  function  $g(\hat x,x)$, which quantifies the gain [or the loss, when the value of $g(\hat x, x)$ is negative] obtained by estimating $\hat x$ when the actual value is $x$. 
In the following, we will require that the payoff function is \emph{positive}, that is, 
\begin{equation}
g(\hat x, x)\ge 0, \quad \forall \hat x,x \in \mathsf X.
\end{equation} 
Clearly, this assumption can be made without loss of generality  as long as the payoff is lower bounded  (that is, as long as there is a limit to the losses). 
  
The expected payoff, averaged over the possible true values, is then given by 
\begin{eqnarray}
\nonumber \gamma  \left[  \mathbf T \right] & := \sum_{x \in \mathsf X}  \pi(x)    \sum_{\hat x \in x}   ~    g(\hat x,x) ~ p(  \hat x|  x) \\
\label{defg}&=     \sum_{\hat x \in \mathsf X}  ~      \Tr\left[T_{\hat x}    G_{\hat x} \right] \qquad G_{\hat x}  =   \sum_{x\in \mathsf X} ~\pi(x)   ~   g(\hat x,x) ~  R_{x} .        
\end{eqnarray}

An example of  payoff function is  $g(\hat x, x)  =  \delta_{\hat x, x}$, which gives  a unit gain if and only if  the estimated value $\hat x$ coincides with the true value $x$.  
In this case the average gain  coincides with the average  probability of guessing  the correct value 
\begin{eqnarray}
\nonumber \gamma  \left[  \mathbf T \right]   \equiv p_{succ}  := \sum_{x \in \mathsf X}  \pi(x)    p( x|  x). 
\end{eqnarray}


A tester $\mathbf T$ is optimal if it achieves the maximum payoff, defined as 
\begin{eqnarray}\label{primal}
\nonumber \gamma_{\max}   :  =& 
\max_{  \mathbf T,  \Xi^{(1)} , \dots , \Xi^{(N)}}  \gamma  \left[  \mathbf T \right] &  \\
\nonumber & T_{\hat x}  \ge 0  , \forall \hat x \in \mathsf X &\\
\nonumber   &   \begin{array}{rcl}
  \sum_{\hat x\in\mathsf X}  T_{\hat x}   & =  &  I_{out,s_N}  \otimes \Xi^{(N)}\\
 \Tr_{in,s_N} \left[\Xi^{(N)}\right]  & =&  I_{out,s_{N-1}}  \otimes \Xi^{(N-1)}\\
   & \vdots &  \\
 \Tr_{in,s_1} [ \Xi^{(1)}]& = & 1.
\end{array} 
 &             
\end{eqnarray}  

\subsection{Dual minimization problem}

Maximizing the payoff in Eq. (\ref{primal})  is a semidefinite program.   Using duality theory  we now give a useful expression for the maximum payoff: 
 
 \begin{theo}\label{dualmetro}
The maximum payoff is given by 
\begin{equation}\label{dualopt}
\gamma_{\max}  =  \min \left \{  \lambda    \ge 0  ~|~  \exists  R     \in    \Comb   \left[  \bigotimes_{s\in \mathsf S}   \left (   \spc H_{out}^{(s)} \otimes \spc H_{in}^{(s)}  \right)  \right]    :   \lambda  R  \ge   G_{\hat x} ,\quad \forall {\hat x}  \in\mathsf X  \right\}, \qquad
\end{equation} 
where $G_{\hat x}$ is defined as in Eq. (\ref{defg}).
 \end{theo}

The proof of the theorem, given in the Appendix, follows the same lines used by Gutoski  \cite{gutoskinorm} to prove strong duality for the minimum error discrimination of two quantum processes, which the  special instance of our problem corresponding to $\mathsf X  :  =\{  0,1\}$ and $g(\hat x, x)  =  \delta_{\hat x, x}$.         
Here we illustrate the result of theorem \ref{dualmetro} in a few special examples.

\subsection{Examples}
\subsubsection{State estimation.}   State estimation can be viewed as a special case where the unknown process $ \map P_x $ to be estimated consists only in the preparation of a quantum state $\rho_x  \in \Lin_+  (\spc H)$ (that is, when there is only one time step $N=1$, the output Hilbert space is $\spc H_{out}^{(s_1)}    =  \spc H$, and  the input Hilbert space is trivial  $ \spc H_{in}^{(s_1)} \simeq \Cmplx$ ).   In this case, the expression (\ref{dualopt}) becomes      
\begin{eqnarray}\label{statest} 
\gamma_{\max}  =  \min \left \{  \lambda    \ge 0  ~|~  \exists  \rho   \in \St \left(   \spc H  \right)  :   \lambda \rho   \ge   G_{\hat x} ,\quad \forall \hat x\in\mathsf X  \right\} ,
\end{eqnarray} 
with $G_{\hat x}  =   \sum_{x\in \mathsf X}  \pi(x)  ~  g(  \hat x, x)  ~ \rho_{ x }$.

\subsubsection{Minimum error state discrimination.}     
If  $g(\hat x, x)  =  \delta_{\hat x, x}$, the maximum payoff $\gamma_{max}$ coincides with the maximum probability of guessing the correct value  $ p^{\max}_{succ}$, so that maximizing the payoff is equivalent to minimizing the error probability.       
In this special case  we retrieve from Eq. (\ref{statest}) the classic expression by Yuen, Kennedy, and Lax \cite{yuen}  (see also \cite{fiura,hayashi})  
\begin{eqnarray}\label{old} 
p^{\max}_{succ}  =  \min \left \{  \Tr[\Lambda]   ~|~  \Lambda \in  \Lin  (\spc H)   , \quad     \Lambda   \ge     \pi_{\hat x}    \rho_{\hat x} , \forall \hat x\in\mathsf X  \right\} 
\end{eqnarray} 
[the above expression follows from Eq. (\ref{statest}) with the definition $\Lambda: =  \lambda  \rho$].

\subsubsection{State estimation/discrimination in the group covariant case.} 
    The dual expression for the maximum payoff has an interesting interpretation in the presence of symmetry.  Let us first consider a simple case of   state discrimination, where   $\mathsf X$  is a finite group,  the prior probability $\pi$  is uniform, that is, $\pi(x)  =  1/ |\mathsf X|$, and the unknown state $\rho_x$ is given by $\rho_x  =  U_x \rho_0  U_x^\dag$, where $\rho_0  \in \St(\spc H) $ is a fixed state and $U :  \mathsf X \to \Lin  (\spc H) , x \mapsto  U_x $  is a projective unitary representation of the group $\mathsf X$.   In this case, it is easy to show that  the minimization over $\Lambda =  \lambda \rho$ in Eq. (\ref{old}) can be restricted without loss of generality to invariant states, satisfying $ U_x \rho U_x^\dag  =  \rho, \forall x\in  \mathsf X $.   Hence, we have
 \begin{eqnarray}\label{psucc}
\nonumber p^{\max}_{succ}  &=&  \min \left \{  \lambda   ~|~  \exists  \rho \in\St (\spc H) :  \rho  {\rm ~is~invariant},  \rho   \ge     \frac  {  \rho_0}{\lambda |\mathsf X|}   \right\} \\
    & =&   \frac  1    {   |\mathsf X|  q_{\max} }  \\
 \nonumber & &  q_{\max}  :  =    \max \left \{  q  ~|~    \exists  \rho \in\St (\spc H) :  \rho  {\rm ~is~invariant} ,    q  \rho_0  \le \rho   \right\}  
\end{eqnarray} 
By definition, $q_{\max}$ is the maximum probability that  $\rho_0$  can have in an ensemble decomposition of an invariant state $\rho$, optimized over all possible invariant states.  The probability $q_{\max}$ ranges between  $1/|\mathsf X|$ and $1$. Intuitively, $q_{\max}$ can be interpreted as a measure of how symmetric the state $\rho_0$ is: for $q_{\max}=1$ the state $\rho_0$ is invariant, while for $q_{\max}  = 1/|\mathsf X|$  the state $\rho_0$ generates a family of orthogonal states $ \rho_x =  U_x  \rho_0  U_x^\dag$.   

The result can be easily extended to the case of arbitrary payoff functions that are \emph{left-invariant} under the action of the group, that is, functions $g$ satisfying the condition $g(y  \hat x,  y x)  =  g(\hat x, x),  \forall \hat x,  x, y\in \mathsf X$. Moreover,  the expression of Eq. (\ref{psucc}) can be generalized to a form that holds also for continuous groups:    

\begin{cor}\label{cor:uno}
Let $\mathsf X$ be a compact group, $g :  \mathsf X  \times \mathsf X  \to \Reals$ be a left-invariant payoff function, and $\rho_x$ be the quantum state $\rho_x  : =   U_x \rho_0 U_x^\dag$, where $U:  x \mapsto U_x $ is a unitary representation of the group $\mathsf X$.   
If the prior probability is given by the Haar measure  $\d x$, then the maximum average payoff over all quantum measurements is given by     
 \begin{eqnarray*}
 \gamma_{\max}  =   \frac{  \gamma_0 } {q_{\max}} \qquad  & \gamma_0  :  =   \int_{\mathsf X} \,  {\rm d} x ~  g(e, x) \\
 &  q_{\max} : =   \max \left \{  q  ~|~    \exists  \rho \in\St (\spc H) :  \rho  {\rm ~is~invariant} ,    q  \sigma_0  \le \rho   \right\} \\
 &  \sigma_0  : =  \frac 1 {\gamma_0}  \int_{\mathsf X}   {\rm d} x ~   g  (e, x) ~  U_x  \rho_0  U_x^\dag  ,            
 \end{eqnarray*} 
where $e\in\mathsf X$ denotes the identity element in the group $\mathsf X$.  
\end{cor}

\Proof Using the invariance of the Haar measure and of the payoff function it is easy to check that $G_{\hat x}  =    U_{\hat x} (\gamma_0  \sigma_0)  U_{\hat x}^\dag$.    Using this fact, we can restrict the minimization  in Eq. (\ref{statest}) to invariant states $\rho$ satisfying the condition $ \lambda \rho   \ge  \gamma_0  \sigma_0$.    Finally, defining $q : =    \gamma_0/\lambda$ we can transform the minimization over $\lambda$ into a maximization over $q$, thus proving the thesis. \qed 

\subsubsection{Binary discrimination of multi-time quantum processes}  

The discrimination of two multi-time processes $\map P_0$ and $\map P_1$  corresponds to the special case where $\mathsf X  =  \{0,1\}$.  In this case, the maximum probability of successful discrimination defines an operational norm in the real vector space  generated by quantum processes    \cite{memorydisc,gutoskinorm}.  
For prior probabilities $\pi_0$ and $\pi_1$,  the probability of success and the norm are linked by the relation  \cite{memorydisc}
\begin{equation*}
 p_{succ}   =    \frac 12   \left(  1 +   |\!|  \pi_0  \map P_0  -  \pi_1 \map P_1   |\!|_{op}  \right) ,     
\end{equation*}     
which generalizes the well-known expression by Helstrom \cite{hel} for the optimal discrimination between two quantum states.  In the binary case the dual expression for the maximum success probability given by theorem \ref{dualmetro} coincides with the dual expression presented by Gutoski in  Ref. \cite{gutoskinorm}.

\subsubsection{Process estimation/discrimination in the group covariant case.}
Consider the case of a general process $\map P_x$ consisting of $N$ time steps. Suppose that $\map P_x$  has the form  $\map P_x  =  \left(  \bigotimes_{ s  \in  \mathsf S} \map V^{(s)}_x \right)   \map P_0   \left(  \bigotimes_{ s  \in  \mathsf S} \map U^{(s)  \dag}_x \right)  $, where $\map P_0$  is a fixed process and $\map U^{(s) \dag}_x (\rho)  :=   U_x^{(s) \dag}  \rho  U^{(s) }_x$  $\left[ \map V^{(s)}_x (\rho)  := V_x^{(s)}  \rho  V^{(s)  \dag}_x \right]$ is a unitary quantum channel representing the action of the group on the input (output) system at the $s$-th time step.        

Denoting by $R_x$ and $R_0$ the quantum combs corresponding to the processes $\map P_x$ and $\map P_0$, it is possible to show  that $R_x =   \left (   \bigotimes_{s\in\mathsf S}    \map V^{(s)}_x \otimes \map U_x^{(s)   *}   \right)  (R_0) $  where $ \map U_x^{(s) *} $ denotes the complex conjugate $\map U_x^{(s)  *}$ with respect the computational basis \cite{covcomb}.    

The result of Corollary \ref{cor:uno}  can then be generalized immediately to the case of general processes: 

\begin{cor}
Let $\mathsf X$ be a compact group, $g :  \mathsf X  \times \mathsf X  \to \Reals$ be a left-invariant payoff function, and let $\rho_x$ be the quantum state $\rho_x  : =   U_x \rho_0 U_x^\dag$, where $U:  x \mapsto U_x $ is a unitary representation of the group $\mathsf X$.   
If the prior probability is given by the Haar measure  $\d x$, then the maximum average payoff over all quantum measurements is given by     
 \begin{eqnarray*}
 \gamma_{\max} & =   \frac{  \gamma_0 } {q_{\max}}  \\
  \gamma_0    &:  =   \int_{\mathsf X}  {\rm d} x ~  g(e, x) \\
  q_{\max} &: =   \max \left \{  q  ~|~    \exists  R \in \Comb   \left (   \bigotimes_{s\in\mathsf S}    \spc H^{(s)}_{out} \otimes \spc H^{(s)}_{in}    \right)   :   R  {\rm ~is~invariant} ,    q  S_0  \le R   \right\} \\
   S_0  &: =  \frac 1 {\gamma_0}  \int_{\mathsf X}   {\rm d} x ~   g  (e, x) ~      \left (   \bigotimes_{s\in\mathsf S}    \map V^{(s)}_x \otimes \map U_x^{(s) *  }   \right)  (R_0)    ,            
 \end{eqnarray*} 
where $e\in\mathsf X$ denotes the identity element in the group $\mathsf X$.  
\end{cor}

\Proof Same proof as for corollary 1. \qed  
 \section{Product rule for the estimation of independent processes}

Imagine that we have $K$  processes, where each process $\map P_{k,x_k}$ corresponds to a quantum network as in figure (\ref{withmemo}) and is labelled by an unknown parameter $x_k$ in some set $\mathsf X_k$, $k=1,\dots, K$.   For every fixed $k$, all the processes $\{\map P_{k,x_k}~|~   x_k  \in \mathsf X_k  \}$  consist of  the same number $N_k$ of time steps, which we label  by an index $s_k$ in some set $ \mathsf S_k  \subset \mathbb N$.      At time $s_k$, each process  $\map P_{k,x_k}$  will transform an input system with Hilbert space   $ \spc H^{(s_k)}_{k,in} $, into an output system  with Hilbert space $\spc H^{(s_k)}_{k,out}$.   \newline

Let us denote by   ${\bf x}$ the vectors of parameters  ${\bf x} : =   (x_1, \dots, x_K)\in  \mathsf X  :=  \mathsf X_1 \times \cdots \times \mathsf X_K  $.    
We say that  the $K$ processes $\{ \map P_{k,x_k} ~|~  k=1,  \dots , K\}$  are \emph{independent} when
\begin{itemize}
\item  two processes  $\map P_{k, x_k}  =    \map P_{l, x_l}$ with $k\not = l$  correspond to two disconnected quantum networks for every $x_k \in\mathsf X_k$ and for every $x_l\in\mathsf X_l$
\item  the prior distribution of the parameters factorizes as 
\begin{eqnarray}
\pi({\bf x})  =  \pi_1  (x_1)  \pi_2  (x_i)  \cdots  \pi_K (x_K), 
\end{eqnarray}
where $\pi_k$ is the prior distribution for the parameter $x_k$.     
\end{itemize}
 
 For example, the different parameters could be $K$ independent and uniformly distributed phase shifts.   

If  $\{ \map P_{k,x_k} ~|~  k=1,  \dots , K\}$  are $K$ independent processes, we denote by  
 $\map P_{\bf x}  :=   \map P_{1, x_1}   \otimes  \map P_{2,x_2}  \otimes \cdots \otimes \map P_{K,x_K}$ the corresponding joint process.

Suppose that we want to estimate parameter ${\bf x}$ labelling the joint process  $\map P_{\bf x}$  and that our figure of merit is given by the payoff function  $g( \hat{\bf x}, \bf x)$.   If we are interested in each parameter independently, then  the payoff function for the estimation of the vector $\bf x$ is the product of the payoff functions for the estimation of its components:  
\begin{eqnarray}\label{prodpayoff}      
g ( \hat{\bf x}, {\bf x}) =  \prod_{k=1}^K   g_k  (\hat x_k, x_k)   \qquad  g_k   \ge 0  ,\forall k = 1,\dots, K,
\end{eqnarray}
where the notation $g_k\ge 0$ means $    g(\hat x_k , x_k)  \ge 0,  \forall  \hat x_k , x_k  \in  \mathsf X_k$.
 For example, the payoff function could give a unit reward only when all the parameters are guessed correctly, so that $g(\hat {\bf x}, {\bf x}) =  \delta_{\hat {\bf x}, {\bf x}} =\prod_{k=1}^K  \delta_{\hat x_k, x_k}$.
  
Note that, in order to have a meaningful figure of merit for the estimation of the vector $\bf x$, it is important to have $g_n  \ge 0 $ for every $n$: otherwise, the product of two negative gains (i.e. of two losses) for two different parameters would count as a positive gain for the joint estimation of the vector $\bf x$.

Based on the hypotheses of independence of the processes and on the product form of the payoff function we can prove the following theorem:  

\begin{theo}\label{theo:prod} {\bf (Product rule for the estimation of $K$ independent processes)}
Let $\map P_{k, x_k}  , k=1, \dots, K$ be $K$ independent processes, each process labelled by an unknown parameter $x_k \in \mathsf X_k$  with prior probability $\pi_k  (x_k)$.    Then for a payoff function $g(\hat {\bf x}, {\bf x})$  of the product form of Eq. (\ref{prodpayoff})  the maximum payoff for the estimation of $\bf x$  is given by the product of the maximum payoffs for the the estimation of its components:  
\begin{equation}
\gamma_{\max}      =    \prod_{k=1}^K   \gamma_{\max, k},     
\end{equation}       
where $\gamma_{\max,k}$  is the maximum payoff achievable in the estimation of $x_k$.  

In other words, the optimal estimation of the vector $\bf x$ can be achieved by estimating each component $x_k$ independently.
\end{theo}  

\Proof  Clearly, we have $\gamma_{\max}      \ge    \prod_{k=1}^K   \gamma_{\max, k}$, because restricting to product strategies can only reduce the maximum payoff.
To prove the converse we use the dual minimization problem  of Theorem \ref{dualmetro}, in which restricting  to product combs can only increase the minimum.    
   
Let $ R_{k, x_k}$ be the quantum comb representing the process     $\map P_{k,x_k} $ and let $  R_{\bf x}   =   \bigotimes_{k=1}^K   R_{k,x_k}  $ be the quantum comb representing the process $\map P_{\bf x}  =   \bigotimes_{k=1}^K  \map P_{k, x_k} $.    Let us introduce the notation   
\begin{eqnarray*}
\mathsf C_k & :  =  \Comb \left[  \left(   \bigotimes_{s_k \in\mathsf S_k}   \spc H^{(s_k)}_{out}  \otimes  \spc H^{(s_k)}_{in} \right)  \right]\\
\mathsf C & :  =  \Comb \left[  \left( \bigotimes_{k=1}^K   \bigotimes_{s_k \in\mathsf S_k}   \spc H^{(s_k)}_{out}  \otimes  \spc H^{(s_k)}_{in} \right)  \right]\\
\mathsf C_{prod}  &: =  \left \{   R  =  \bigotimes_{k=1}^K   R_k  ~|~   R_k \in \mathsf C_k ~ \forall k=1,\dots, K \right \}  \subset \mathsf C.  
\end{eqnarray*}
With this notation we have  that $R_{k, x_k}$ and $R_{\bf x}$ belong to $\mathsf C_k$ and $\mathsf C$, respectively. 

Define the positive operators 
\begin{eqnarray*}
G_{k,  \hat x_k}  & :  =   \sum_{x_k \in  \mathsf X_k}     \pi_k  (x_k)  ~  g_k (\hat x_k,x_k) ~ R_{k, x_k}  \\
G_{ \hat {\bf x}}    & :  =   \sum_{ {\bf x} \in  \mathsf X}     \pi  ({\bf x})  ~  g ( \hat {\bf x},  {\bf x})  ~  R_{{\bf x}}   \equiv  \bigotimes_{ k=1 }^K  G_{k,  \hat x_k}  .                
\end{eqnarray*}
Then, by theorem \ref{dualmetro} we have 
\begin{eqnarray*}
\gamma_{\max} & =  \min \left \{  \lambda    \ge 0  ~|~  \exists  R     \in \mathsf C   :   \lambda  R    \ge   G_{\bf x} ,\quad \forall {\bf x}  \in\mathsf X  \right\}\\  
                            & \le \min \left \{  \lambda    \ge 0  ~|~  \exists R     \in \mathsf C_{prod}    :      \lambda R   \ge   G_{\bf x} ,\quad \forall {\bf x}  \in\mathsf X  \right\}\\
                       & \le \prod_{k=1}^K   \min \left \{  \lambda_k    \ge 0  ~|~  \exists R_k     \in \mathsf C_k   :  \lambda_k   R_k     \ge   G_{k, x_k} ,\quad \forall {x_k}  \in \mathsf X_k  \right\}    \\
                        & = \prod_{k=1}^K  \gamma_{\max,k}.              
\end{eqnarray*} 
Here, the second inequality comes from the fact that    if $ \lambda_k  R_k  \ge  G_{k,  x_k}$ for all $k$, then    $\lambda R  \ge G_{\bf x}$   for $\lambda =  \prod_k  \lambda_k$  and $R  =  \bigotimes_k R_k $.   \qed 

\subsubsection{Relation with the product rules by Mittal and Szegedy.} The technique used to prove that the optimal payoff is of the product form is directly inspired by a result by Mittal and Szegedy on product rules for semidefinite programming \cite{mittszeg}.  However, our result is not a  direct application of the theorem in Ref. \cite{mittszeg}, which concerns \emph{product programs}, where the linear constraint for the product program is the tensor product of the linear constraints for the individual programs.  The theorem is not directly applicable in our case because in the joint estimation of $K$ processes the linear constraint of Eq. (\ref{primal})  are not the tensor product of the linear constraints for the estimation each process separately.   However, the crucial point here is that the tensor product  of $K$ operators satisfying the constraints individually is an operator that satisfies the joint constraint and that this property is true both in the primal maximization problem and in the dual minimization program.  

\subsubsection{Example 5: minimum error discrimination of $K$ sets of processes}   Theorem \ref{theo:prod} can be applied to the case of minimum error discrimination of processes.    Suppose that for every $k=1,\dots, K$ we have a set of processes  $\{  \map P_{k_, x_k }  ~|~  x_k  \in  \mathsf X_k    \}$,  each process $\map P_{k_, x_k }$ having prior probability $\pi_{k,x_k}$  ($\sum_{x_k \in\mathsf X_k} \pi_{k,x_k}  =1  $).     Denoting by $p^{\max}_{succ, k}$ the maximum probability of success in correctly identifying the $k$-the process,  and by $  p^{\max}_{succ} $ the probability of success in correctly identifying all processes, we then have $
p^{\max}  =  p^{\max}_{succ, 1 }  \cdots  p^{\max}_{succ, K }$. The best joint strategy for discrimination is just the product of the best individual strategies.     

\subsection{Counterexamples }
Our theorem \ref{theo:prod} proved the optimality of product strategies in the hypotheses that the processes are independent \emph{and} that the payoff function is of a product form.  Here we show that if one of these hypotheses is dropped, there are examples where the result  does not hold.   

\subsubsection{Minimum error  discrimination of two pure states with multiple copies.}     
One of the most basic problems in quantum information is to distinguish between two non-orthogonal quantum states (see e.g. the classic textbook of Helstrom \cite{hel}).  In this context, one important question is how small the probability of error can be made when a finite number of identically prepared quantum systems  are available.     Consider the minimum error discrimination of two pure states $\{ \rho_0,  \rho_1   \}$ with prior probabilities $\{p_0, p_1\}$, in the case    where  $K$ identical copies of the unknown state are available.  We can view this problem as an instance of minimum error discrimination of $K$ perfectly correlated preparation processes, each of which prepares one of the states $ \{\rho_0 , \rho_1 \}$.     Denoting by $p^{\max}_{succ} (K)$ the probability of success with $K$ copies, we know from the quantum Chernoff bound \cite{chernoff} that $p^{\max }_{succ} (K) $ converges to 1 exponentially fast in the limit  $K\to \infty$.  On the other hand, the product of the probabilities of success, given by   $  \left[p^{\max }_{succ} (K=1)\right]^K$ tends to zero (exponentially fast) unless the two states are perfectly distinguishable.

\subsubsection{Estimation of two independent phase shifts with a correlated payoff function.}  

Phase estimation is another great classic of quantum estimation theory \cite{hel,hol}, with  applications to quantum clocks \cite{phase} and high-precision interferometry (see \cite{qmetro1,qmetro2} for an overview of the relevant literature).   In the usual scenario, one has given access to multiple queries to the same black box implementing an unknown phase shift and the question is how the precision of estimation increases with the number of queries \cite{phase,phasenet}.     Here we will consider instead a different scenario:  two black boxes implementing different (uncorrelated) phase shifts are given and the goal is to estimate the values of the two shifts.  A priori, since the the values of the two phase shifts are independent, it could sound natural that the optimal estimation strategy consists in estimating each phase shift independently.   However, in the following we will see that an arbitrarily small amount of correlation \emph{in the figure of merit}  used to judge the quality of the estimation can change critically the features of the optimal network, with the optimal input state changing suddenly from factorized to maximally entangled.    

Let us see  in detail how the example works. Consider the estimation of two independent phase shifts on two qubit systems, with Hilbert spaces $\spc H_1$ and $\spc H_2$, respectively ($\spc H_1  \simeq \spc H_2 \simeq \Cmplx^2$).    Denoting by $|0\>$ and $|1\>$ the two orthonormal vectors in the standard basis for $\Cmplx^2$, the phase shifts on a qubit system are given by $U_x  =   |0\>\<0|  +   e^{i x}  |1\>\<1|$, $x\in  [0, 2\pi)$. We assume that the phase shifts on the two qubits are uniformly distributed according to the Haar measure  $\d x  /2\pi$.   The problem is then to find the best estimate of the unknown parameter ${\bf x} :  = (x_1, x_2)$  characterizing the black boxes $U_{x_1}$ and $U_{x_2}$.   
As a figure of merit, we consider the maximization of the payoff function  
\begin{eqnarray*}
g_p(\hat {\bf x} , {\bf x} )  =      p   \cos ( \hat  x_1  +\hat  x_2 - x_1 - x_2  )   + (1-p)     \cos ( \hat x_1 -  \hat x_2 -  x_1 +  x_2),  
\end{eqnarray*}
for some $p  \in  [0,1]$.  
Note that $g_p$ is a convex combination of the figure of merit  $ \cos ( \hat x_1  + \hat x_2 -  x_1 -  x_2 )$, which quantifies how good is our estimate of the sum  $s:=x_1  +  x_2$, and of the figure of merit $ \cos ( \hat x_1  -  \hat x_2 -  x_1 +  x_2 )$, which quantifies how good is our estimate of  the difference $d: = x_1-  x_2$.   In other words, we can interpret $f$ as expressing the fact that, with probability $p$, we will be asked to estimate the sum, while with probability $(1-p)$ we will be asked to estimate the difference.   

Due to the symmetry of the problem, is is enough to consider quantum networks where the two unknown phase shifts are applied in parallel   on a suitable entangled state  $|E\>  \in  \spc H_1 \otimes \spc H_2$, as proven in Ref. \cite{memorydisc}. No additional reference system is needed, because the black boxes form a unitary representation of an Abelian group \cite{entest}.    Hence, the problem is reduced  to the optimal estimation of $\bf x$  from   from the output  state 
$|E_{\bf x}  \>  :=  (U_{x_1} \otimes U_{x_2}) |E\>$.

From the theory of optimal estimation of group parameters \cite{entest} we know that the optimal measurement is given by the covariant POVM 
\begin{eqnarray*} 
 P_{\hat {\bf x}}  =   (U_{x_1} \otimes U_{x_2}) |\eta\> \< \eta |    (U_{x_1} \otimes U_{x_2}) ^\dag  \qquad |\eta\>  := |0\>|0\>  + |0\>|1\>  + |1\>|0\>  + |1\>|1\> .
 \end{eqnarray*}  
Incidentally, we note  that the POVM is of the product form $P_{\hat {\bf x}}  =  P_{1, \hat x_1}  \otimes P_{2, \hat x_2}$.   
By direct calculation, we then find that the average value of $g_p$ is $\gamma_p  =   \<  E  |   G_p  |E\> $ with  
\begin{eqnarray*}
 G_p  =  \frac p 2  ( |0\>|0\>  \<1|  \< 1|  +   |1\>|1\>  \<0|  \< 0|  )  +  \frac{1-p} 2  ( |0\>|1\>  \<1|  \< 0| +  |1\>|0\>  \<0|  \< 1|)   .
 \end{eqnarray*}
  Clearly,  the maximum eigenvalue of $G_p$  is $\lambda_{\max}  =  \max\{  p/2, (1-p)/2\}$, corresponding to the nondegenerate eigenvector 
$|E\>  =  2^{- \frac 12}  (|0\>|0\>  +   |1\>|1\>) $ for $p > 1/2$  and  $|E\>  =  2^{- \frac 12}  (|0\>|1\>  +   |1\>|0\>) $ for $p < 1/2$.   
For $p  =1/2$ one has degeneration, and the optimal input state  can be chosen of the product form  $|E\>  =  | +\>  |+\>$ with $|+  \>   =  2^{-\frac 12} ( |0\>  + |1\>) $.  

The qualitative explanation of the behaviour is the following:      For $p= 1/2$ the figure of merit is factorized ($g_{\frac 12} =  \cos (\hat \varphi  -  \varphi)   \cos(\hat \psi  -  \psi) $) and  the optimal estimation strategy can be chosen to be factorized too.    For every value  $p  \not = \frac 1 2$, the degeneration is removed and suddenly  the optimal input state becomes maximally entangled.  The optimal input state depends in a discontinuous way from the parameter $p$:  the (unique) optimal input state for $p>1/2$  is orthogonal to the (unique) optimal input state for $p<1/2$.    Note, however, that there is no discontinuity in the average payoff.

\subsubsection{Estimating the sum of $K$ independent phase shifts. }   The relation between the correlations in the figure of merit and the correlations in the optimal estimating network can also be observed in the case  of multiple independent phase shifts.    
Suppose that we have $K$ identical systems, with Hilbert spaces $ \spc H_k \simeq  \Cmplx^{N}$ for all $k=1,\dots, K$, and   suppose that each  system undergoes an independent phase shift   $U^{(k)}_{x_k}  :  =    e^{ i  x_k H^{(k)}}$, where $H^{(k)}  :=   \sum_{n=1}^N    n  ~ |n\>\< n|$ for every $k$, $\{|n\>\}$ being the computational basis.   

If we want to estimate the sum  $ s: =  \sum_{k}  x_k  $   a  natural  figure of merit is the minimization of the expected value of the cost function
$c(\hat s , s  ) =    2   [ 1  -\cos(  \hat s - s)   ]$.   
This cost function is well known in the phase estimation literature as a smooth and periodic version of the variance \cite{hel,hol,phase,phasenet}.  For small $s$, we have indeed $ \hat c(\hat s, s)   \approx   ( \hat s - s)^2    $.   Clearly, minimizing $c$ is equivalent to maximizing the payoff function $g (\hat s, s)  = 1+  \cos (  \hat s -s)$.  

Let us find the optimal estimation strategy.  First, using the fact that the unknown black boxes form a unitary representation of an abelian group, we know that the optimal strategy consists in applying the black boxes in parallel on an entangled input state $|E\>  \in  \spc H^{\otimes K}$ \cite{memorydisc,entest}.  
 Moreover, note that for every fixed $i$ and $j$, if we apply the transformation $x_i  \mapsto  x_i  +  \xi $,  $\hat x_i  \mapsto  \hat x_i  +  \xi $,  $x_j \mapsto  x_j -  \xi $,  $\hat x_j  \mapsto \hat  x_j  -  \xi $,  $\xi\in[0,2\pi)$, then the value of the figure of merit does not change.  Using this symmetry it is easy to show that the input state $|E\>$ must be an eigenstate of the difference operator $\Delta_{ij}  = H^{(i)}  -  H^{(j)}$ for every possible pair $i,j$.   It is then straightforward that the optimal choice is $|E\> = \sum_{n=1}^N   e_n  |n\>^{\otimes K} $, where $\{e_n\}$ are suitable coefficients.  The problem then becomes to estimate the sum $s$ from the state $|E_{\bf x}\>   :  =  \left( \prod_k U^{(k)}_{x_k} \right) |E\rangle  = \sum_{n=1}^N    e^{is n} e_n  |n\>^{\otimes K}$.   From the theory of optimal phase estimation we know that  the minimum cost is  $c_{\min}  =4 \sin^2\left[\frac{\pi}{2  N}\right] $, which converges to $ \frac{\pi^2  }  {N^2}$ in the limit $N \to \infty$ (see  Ref. \cite{phase}).   The corresponding optimal state is the entangled state \cite{phase}
 \begin{eqnarray*} 
 |E_{opt}\>   =      \left( \frac{N}2\right)^{-\frac 12}      \sum_{n=1}^{N} \sin\left[\frac{\pi (n-1)}{(N-1)}\right] 
  |n\>^{\otimes K}  .  
 \end{eqnarray*}      
and the optimal POVM is  $P_s   = |\eta_s\>\<  \eta_s  |$, $|\eta_s\>  :  =  \sum_{n=1}^N   e^{isn}  |n\>^{\otimes K}$.    
 It is easy to see that the use of entanglement implies an advantage over factorized strategies, where each system is prepared independently in a state $|e_k\>$ and is measured independently with the optimal POVM.   Indeed, if we choose the optimal states 
$|e_k\>    = |e\>   :=   \left( \frac{N}2\right)^{- \frac 12}      \sum_{n=1}^{N} \sin\left[\frac{\pi (n-1)}{N-1}\right]$ and the optimal product POVM $P_{\hat {\bf x}}  :=  \prod_k     U^{(k)}_{x_k}   (2 |+\>\<+|) U^{(k)\dag}_{x_k} $ then  
we obtain the cost 
\begin{eqnarray*}  
\<  c (\hat s, s)\>  &  =  2  (1  -  \<   \cos (\hat s-s)  \>   )  \\
&  =     2  \left (1  -  \prod_{k=1}^K \<    \cos (\hat x_k-x_k)  \>   \right)   \\
  &   =  2 \left\{ 1-  \left[  1-  2 \sin^2\left(\frac{\pi}{2  M}  \right)\right]^K \right\},
\end{eqnarray*}
where $\<   f \>$ denotes the expectation value of the function $f$.
For large $N$ we get  the asymptotic expression     $\<c\>  \approx  \frac{K\pi^2}{N^2}$. From the comparison with the optimal value $c_{\min}  \approx \frac{\pi^2}{N^2} $ we note that entangling $K$ systems and performing a joint measurement implies a reduction of the variance of a factor $K$ in the estimation of the sum.

\section{Conclusions}  

In this paper we addressed the estimation of an unknown quantum process that can possibly consist of a finite number of time steps.  
 We formulated the search of the optimal quantum network for estimation as a semidefinite program, and used duality theory to give an alternative expression of the maximum payoff achieved by the optimal network.   Using this result we proved a product rule for quantum metrology, showing that the individual strategies are sufficient to achieve the optimal joint estimate of a set of independent processes whenever the figure of merit is of the product form.  In particular, the probability of success in the discrimination of $K$ sets of processes is the product of the probabilities of success for each set.  
 
It is easy to see that the product rule established here for joint estimation can also be extended to the optimization of quantum networks for other tasks, such as the optimal cloning of independent sets of states and processes.  In the case of pure state cloning, it has been observed in Ref. \cite{money}  that the product rule shows that the maximum global fidelity for the joint cloning of $K$ sets of states is the product of the maximum global fidelities for each set, so that the optimal joint cloner is the product of the optimal individual cloners. Using the same type of argument, one can show  that the global channel fidelity for the  joint cloning of $K$ sets of unitary gates (see Ref. \cite{clonunit} for the definition of the cloning task) is the product of the maximum global fidelities for each set, so that the optimal joint cloning network is the product of the optimal individual networks.

\bigskip 
{\bf Acknowledgements.} 
This work is supported  the National Basic Research Program of China (973) 2011CBA00300 (2011CBA00301).
The author gratefully acknowledges the hospitality of the Institute of Theoretical Computer Science and Communications, Chinese University of Hong Kong, where this work has been completed.   A particular thanks goes to G Gutoski for pointing out the proof of strong duality in Ref. \cite{gutoskinorm} and to the referee for useful observation that helped improving the presentation.      

\section*{Appendix}

{\bf Proof of theorem \ref{dualmetro}.}   Define the block diagonal matrices $T : =  \left(   \bigoplus_{n=1}^N   \Xi^{(n)}  \right)  \oplus \left(   \bigoplus_{x\in\mathsf X}    T^{(N)}_x  \right) $   and $G  =   \left(   \bigoplus_{n=1}^N   0_n  \right)  \oplus \left(   \bigoplus_{x\in\mathsf X}    G^{(N)}_x  \right)   $, where $0_i$ denotes the zero matrix in the $i$-th block.
With these definitions, the optimization problem in Eq. (\ref{primal})  can be written as a semidefinite program in the standard form 
\begin{equation}
\begin{array}{llr}
 \gamma_{\max}& =  \max_T  ~  & \Tr[  T  G ]   \\  
&\nonumber {\rm subject~to} ~&  T \ge 0\\
&\nonumber & \map L  (T)  =  K          
\end{array}   
\end{equation}
where $\map L$ is the Hermitian-preserving linear map defined by  $\map L ( T )  = \bigoplus_{j=0}^{N}   R^{(j)}$ with  
\begin{eqnarray*}
R^{(0)}  &= & \Tr_{in, s_1} [ \Xi^{(1)}]\\
R^{(1)}  &=&  \Tr_{in,s_2}  [\Xi^{(2)}]  -  I_{out,s_1}  \otimes  \Xi^{(1)}\\
&  \vdots &\\
R^{(N-1)}   &=& \Tr_{in, s_{N}}    [\Xi^{(N)}]  -  I_{out,s_{N-1}}  \otimes  \Xi^{(N-1)}\\
R^{(N)}  &=&\left (  \sum_{x\in  \mathsf X} T_x  \right)- I_{out, s_N}  \otimes \Xi^{(N)}, 
 \end{eqnarray*}  
 and $K$ is the block diagonal operator $  K :  =  \bigoplus_{j=0}^{N}  K^{(j)}$ defined by $K^{(0)}  = 1$  and $K^{(j)}  = 0_j$ for every $j=1, \dots, N$.   

Using the duality of semidefinite programming we obtain 
\begin{eqnarray}\label{canonicaldual}
 \gamma_{\max}  \le  \gamma^*& :=  \min_S  ~  & \Tr[  S  K ]   \\  
&\nonumber {\rm subject~to} ~ & \map L^\dag  (S)  \ge  G     ,    
\end{eqnarray}

where $S  =  \bigoplus_{j=0}^N  S^{(j)}$ and 
$\map L^\dag$ is the dual map defined by $\<  S, \map L ( T)  \>  =  \<  \map L^\dag (S),  T\> $ with $\<S, T\>  :  =  \Tr[S^\dag T]$ is the Hilbert-Schimdt product. 
Using the definition of $\map L^\dag$, it is easy to check that  $\map L^\dag (S)  = \left(   \bigoplus_{n=1}^N   M_n    \right)  \oplus \left(   \bigoplus_{x\in\mathsf X}    M_x  \right) $    where  
\begin{eqnarray*}
M_1  &= &  I_{in,s_1}   S^{(0)} -  \Tr_{out,s_1}[  S^{(1)} ]    \\
M_2  &=&    I_{in,s_2} \otimes S^{(1)} -   \Tr_{out,s_2}  [S^{(2)}]\\
&  \vdots &\\
M_N   &=&     I_{in, s_N}  \otimes  S^{(N-1)}  -  \Tr_{out, s_N}  [S^{(N)}]    \\
M_x  &=&  S^{(N)} \qquad \forall x\in \mathsf X 
 \end{eqnarray*}  
Recalling the definition of $K$ and $G$, the expression for $\gamma^*$  becomes
\begin{equation}
\begin{array}{lcl}
 \gamma^*& =  \min_S  ~&    S^{(0)}   \\  
&\nonumber {\rm subject~to} ~&    I_{in, s_1}   S^{(0)}  \ge \Tr_{out,s_1}[  S^{(1)} ]  \\
&\nonumber \phantom{\rm subject~to} ~&    I_{in, s_2} \otimes S^{(1)} \ge   \Tr_{out, s_2}  [S^{(2)}]  \\   
& \nonumber \phantom{\rm subject~to} ~ &   \vdots \\
& \nonumber \phantom{\rm subject~to} ~  &  I_{in,s_N}  \otimes  S^{(N-1)}  \ge  \Tr_{out,s_N}  [S^{(N)}]     \\  
& \nonumber \phantom{\rm subject~to} ~   &   S^{(N)}  \ge G^{(N)}_x  \qquad\forall x\in  \mathsf X.       
\end{array}   
\end{equation}
Note that $S^{(N)}$ must be positive, since we have $S^{(N)}  \ge G_x^{(N)}  \ge 0$.   Consequently, $S^{(j)}$ must be positive for every $j=0, \dots, N$.    Moreover, there exists at least an operator $S$ such that $  \map L^\dag (S)  > G $.   For example, one can choose  
\begin{eqnarray*}
S^{(N)}  &= & g_{\max} ~   \prod_{n=1}^N   \left (  I_{out,s_n} \otimes I_{in,s_n}  \right)   \qquad g_{\max}  :  = \max_{\hat x, x \in  \mathsf X} g (\hat x, x) \\
S^{(N-1)}  &= &  2~ \Tr_{out,s_N}  \Tr_{in,s_N}  [ S^{(N)} ] \\
&~~\vdots &  \\
S^{(0)} & = &   2~\Tr_{out,s_1}  \Tr_{in,s_1} [S^{(s_1)}]  .   
\end{eqnarray*}

The existence of an operator $S$ such that $\map L^\dag ( S)  >  G$, along with the fact that the maximum payoff $\gamma_{\max}$ is bounded by $g_{\max}$, implies that the hypotheses of Slater's theorem (see e.g. \cite{gutoskinorm,molina}) on strong duality are satisfied.  Hence, the optimum values for the primal and dual optimization problem coincide:  $\gamma_{\max}  =  \gamma^*$.    

Now, we show that the first $N$ inequalities can be chosen to be equalities without loss of generality: we show that for every operator  $S$ satisfying the constraints there exists another operator $\tilde S$ that achieves the equality in the first $N$ constraints and has the same value of the objective function as $S$.  
  To prove this statement, we proceed by induction. First, we define the operator $\tilde S:  =  \sum_{j=0}^N  \tilde S^{(j)}$ through the relations
  \begin{eqnarray*}
  \tilde S^{(0)} & :=S^{(0)}  \\
  \delta^{(1)}  & : =   I_{in, s_1}  \tilde S^{(0)} -  \Tr_{out, s_1}[  S^{(1)} ]  \ge 0  \\
  \tilde S^{(1)}  &:  =    S^{(1)}  + \rho_1 \otimes \delta^{(1)},\\
 \tilde S^{(j)}   &:  =  S^{(j)}  \qquad \forall j = 2,\dots, N   \end{eqnarray*}
where $\rho_1  $ is an arbitrary quantum state in $\St(\spc H_{out, s_1})$.  
 Clearly, with this definition we have $\Tr_{out, s_1}  [\tilde S^{(1)}]  = I_{in, s_1}  \tilde S^{(0)} $, that is, $\tilde S$ achieves the equality in the first constraint.   Moreover, since $\delta^{(0)}$ is positive we have $ I_{in,s_2} \otimes  \tilde S^{(1)}   \ge I_{in,s_2} \otimes   S^{(1)} \ge  \Tr_{out,s_2}[  S^{(2)}] \equiv  \Tr_{out, s_2}[ \tilde  S^{(2)}] $, namely $\tilde S$ satisfies the second constraint.   Hence, the operator $\tilde S$ has the same objective value of $S$, satisfies all the constraints and achieves the equality in the first.       Now, suppose that $S$ achieves the equality in the first $k\ge 1$ constraints and  define 
  \begin{eqnarray*}
  \tilde S^{(j)} & :=S^{(j)}  \qquad \forall j =1,\dots, k  \\
  \delta^{(k+1)}  & : =   I_{in, s_{k+1}}   \tilde S^{(k)} -  \Tr_{out, s_{k+1}}[  S^{(k+1)} ]  \ge 0  \\
  \tilde S^{(k+ 1)}  &:  =    S^{(k+1)}  + \rho_{k+1} \otimes \delta^{(k+1)},\\
 \tilde S^{(j)}   &:  =  S^{(j)}  \qquad \forall j =k+2 ,\dots, N  
  \end{eqnarray*}
where $\rho_{k+1}$ is an arbitrary quantum state in $\St(\spc H_{out, s_{k+1}})$.  With this definition it is immediate to see that $\tilde S$ has  the same objective value of $S$, satisfies all constraints and achieves the equality in the first $k+1$ ones.  By induction, we conclude that for every operator $S$ satisfying the constraints there exists another operator $\tilde S$ which achieves the equality in the first $N$ constraints and has the same objective value.  Defining $\lambda  :  = \tilde  S^{(0)} $  and $R  :  = \tilde S^{(N)}/\lambda$  we then obtain the thesis of the theorem. \qed

\end{document}

%% file: myQcircuit.tex
\usepackage[matrix,frame,arrow]{xy}
\usepackage{iopams}

\newcommand{\qw}[1][-1]{\ar @{-} [0,#1]}

\newcommand{\gate}[1]{*{\xy *+<.6em>{#1};p\save+LU;+RU **\dir{-}\restore\save+RU;+RD **\dir{-}\restore\save+RD;+LD **\dir{-}\restore\POS+LD;+LU **\dir{-}\endxy} \qw}

\newcommand{\multimeasureD}[2]{*+<1em,.9em>{\hphantom{#2}}\save[0,0].[#1,0];p\save !C *{#2},p+LU+<0em,0em>;+RU+<-.8em,0em> **\dir{-}\restore\save +LD;+LU **\dir{-}\restore\save +LD;+RD-<.8em,0em> **\dir{-} \restore\save +RD+<0em,.8em>;+RU-<0em,.8em> **\dir{-} \restore \POS !UR*!UR{\cir<.9em>{r_d}};!DR*!DR{\cir<.9em>{d_l}}\restore \qw}

\newcommand{\multigate}[2]{*+<1em,.9em>{\hphantom{#2}} \qw \POS[0,0].[#1,0];p !C *{#2},p \save+LU;+RU **\dir{-}\restore\save+RU;+RD **\dir{-}\restore\save+RD;+LD **\dir{-}\restore\save+LD;+LU **\dir{-}\restore}
\newcommand{\ghost}[1]{*+<1em,.9em>{\hphantom{#1}} \qw}

\newcommand{\Qcircuit}[1][0em]{\xymatrix @*[o] @*=<#1>}  
 \renewcommand{\Qcircuit}[1][0em]{\xymatrix @*=<#1>}

\newcommand{\pureghost}[1]{*+<1em,.9em>{\hphantom{#1}}}
\newcommand{\multiprepareC}[2]{*+<1em,.9em>{\hphantom{#2}}\save[0,0].[#1,0];p\save !C
  *{#2},p+RU+<0em,0em>;+LU+<+.8em,0em> **\dir{-}\restore\save +RD;+RU **\dir{-}\restore\save
  +RD;+LD+<.8em,0em> **\dir{-} \restore\save +LD+<0em,.8em>;+LU-<0em,.8em> **\dir{-} \restore \POS
  !UL*!UL{\cir<.9em>{u_r}};!DL*!DL{\cir<.9em>{l_u}}\restore}

\newcommand{\poloFantasmaCn}[1]{{{}^{#1}_{\phantom{#1}}}}